# El aula invertida para la docencia de Física
# The flipped classroom for Physics teaching

Patricia Ruiz-Galende[1,2], Mónica Montoya[3], Iciar Pablo-Lerchundi[1], Patricia Almendros[3], Fabio Revuelta[2]

patricia.ruiz@alumnos.upm.es, monica.montoya@upm.es, iciar.depablo@upm.es, p.almendros@upm.es, fabio.revuelta@upm.es

[1]Grupo de Investigación ForPROFE y Grupo de Innovación Educativa Didáctica de la Química, Instituto de la Ciencias de la Educación.

[2]Grupo de Innovación Educativa Física Interactiva y Grupo de Sistemas Complejos, Escuela Técnica Superior de Ingeniería Agronómica, Alimentaria y de Biosistemas

[3]Grupo de Innovación Educativa Química y Análisis Agrícola, Departamento de Química y Tecnología de Alimentos, Escuela Técnica Superior de Ingeniería Agronómica, Alimentaria y de Biosistemas

Universidad Politécnica de Madrid
Madrid, España

*Resumen*- En este trabajo presentamos una experiencia docente on-line con estudiantes universitarios de primer curso basada en la aplicación del aula invertida para el estudio de cinemática de sólido rígido. Los resultados del estudio no demuestran diferencias significativas con respecto a otros temas tratados ni con respecto al mismo tema en el curso anterior, en el que la docencia no fue virtual sino presencial.

*Palabras clave: aula invertida, metodologías activas, Física, sólido rígido*

*Abstract*- In this work, we present an on-line educational experience with first-year university students based on the application of the flipped classroom to study kinematics of rigid solid. Our results show no statistical differences when compared with other parts of the Physics module nor with the results obtained last year, when the educational process was not virtual but in person.

1. INTRODUCCIÓN

La crisis provocada por la COVID-19 ha obligado a cambiar la forma de educar. Ha supuesto que tanto alumnos como profesores deban adaptarse rápidamente a un nuevo modelo como es la enseñanza on-line. En principio puede parecer un modelo atractivo y más interactivo para los estudiantes, pero el escenario en el cual se implanta, la celeridad de su aplicación y la falta de experiencia de la mayoría de los participantes puede llevar , en general, a unos peores resultados (Cifuentes-Faura, 2020). En este sentido, las metodologías activas pueden convertirse en una gran herramienta para la docencia on-line, al permitir que los alumnos sean participantes más activos de su proceso de aprendizaje, incluso en tiempos tan complicados como los que estamos viviendo.

Las metodologías activas incorporan habilidades cognitivas superiores según la taxonomía de Bloom (Santiago, 2019). Sin embargo, muchas veces son difíciles de llevar a cabo debido a la falta de experiencia mencionada anteriormente, a la costumbre por la enseñanza tradicional, y a las complicaciones que suponen el cambio en el rol del docente y la implicación del alumnado (Fidalgo-Blanco et al., 2019; Torres-Belma, 2020). Por ello, se debe hacer el esfuerzo de cambiar el modelo tradicional que fomenta, en cierta manera, la pasividad de los estudiantes, que desarrollan principalmente la memoria, y aprovechar la implantación de la docencia on-line para incorporar este tipo de metodologías, que pueden llegar a estimular el aprendizaje por descubrimiento, el constructivismo y un aprendizaje más significativo y profundo (Murillo, 2007; Torres-Belma, 2020).

Un ejemplo de metodología activa que ayuda a promover la implicación del alumno en su proceso de aprendizaje y que, además, puede ser muy adecuada para la docencia on-line, es el aula invertida ("flipped classroom") (Fidalgo-Blanco et al., 2019). Esta metodología fue popularizada por Jonathan Bergmann y Aaron Sams, quienes encontraron en ella una solución a los problemas de absentismo que observaron en la Escuela Secundaria de Woodland Park en Colorado (EE. UU.) (Bergmann & Sams, 2012). El aula invertida consiste en llevar a cabo las actividades tradicionalmente realizadas en el aula, fuera de ella (sobre todo a través de videolecciones), y dedicar el tiempo de clase a la resolución de dudas y la realización de ejercicios prácticos (Aguilera-Ruiz et al., 2017).

Por todo ello, se escoge el aula invertida como respuesta a la situación provocada por la pandemia, debido a su fácil adaptabilidad al contexto y los estudiantes. Entre sus ventajas destacan las siguientes (Cuevas-Monzonís et al., 2021):

- Flexibilidad en cuanto a la temporalidad y el espacio, ya que los vídeos proporcionados pueden ser visualizados por el alumno dónde y cuándo quiera y tantas veces como necesite, posibilitando una educación asíncrona.

- El aprendizaje queda en manos del estudiante convirtiéndole en agente activo del mismo, lo que favorece un aprendizaje más profundo y significativo (Fidalgo-Blanco et al., 2017).
- El tiempo de clase es mucho más efectivo y el rol docente pasa a ser más el de un mediador y un apoyo para el estudiante.

En este trabajo describimos la implementación del aula invertida para el proceso de enseñanza-aprendizaje de cinemática del sólido rígido (CSR) de la asignatura Física I del Grado en Ingeniería Agroambiental de la Universidad Politécnica de Madrid. Para ello, describimos en la siguiente sección el contexto en el que se ha llevado a cabo la experiencia. A continuación, en la sección 3, presentamos brevemente la metodología usada. En la sección 4 se recogen los principales resultados de nuestro estudio. Finalmente, el artículo se cierra con las conclusiones más importantes de nuestro trabajo.

2. CONTEXTO

La situación pandémica vivida desde marzo de 2020 impidió llevar a cabo la docencia presencial como era costumbre durante buena parte del curso 2019/20 y no ha sido hasta este curso 2020/21 cuando se ha podido ir recuperando paulatinamente la normalidad pre-pandémica. En nuestro centro, no obstante, se resolvió que los estudios de grado continuaran con la docencia on-line también durante todo este curso, por lo que fue necesario adaptar la docencia a la nueva realidad. Por ello, se decidió implementar el aula invertida en el tema 3 de la asignatura de primer curso Física I del Grado en Ingeniería Agroambiental en el que impartimos docencia, que consta del siguiente temario:

1. Cálculo vectorial
2. Cinemática del punto
3. Cinemática del sólido rígido (CSR)
4. Movimiento relativo
5. Estática (equilibrio, centros de gravedad y momentos de inercia)
6. Dinámica de sistemas

Con el objetivo de asegurar el correcto desarrollo de la innovación propuesta, en particular, y de una adecuada docencia on-line, en general, al comienzo del curso se preguntó a los alumnos si contaban con dispositivos adecuados para seguir las clases, respondiendo afirmativamente todos salvo uno (al que se le facilitó un ordenador portátil, dado que seguía las clases con ciertas dificultades usando el móvil y una tableta).

3. DESCRIPCIÓN

Para implementar el aula invertida en el proceso de enseñanza-aprendizaje de CSR, se ha creado una serie de 10 vídeos (con una duración de menos de 5min, como se puede ver en la Tabla 1) a los que los alumnos podían acceder a través de Moodle y que debían visualizar a lo largo de dos semanas en su casa (mientras se trataban en las clases on-line los temas de movimiento relativo y dinámica del punto). En el vídeo inicial se explicaba cómo se iba a llevar a cabo la docencia del tema. Cada uno de los vídeos restantes estaba dedicado a un concepto concreto de CSR, como se detalla en la Tabla 1.

*Tabla 1: Contenidos de los vídeos $V_1$-$V_9$ usados en el aula invertida, junto con su duración ($t_d$), tiempo que tenían los alumnos para responder el cuestionario correspondiente ($t_c$), si lo hubiere, y relación con las preguntas del ejercicio de cinemática del sólido rígido de los exámenes parcial y final durante los cursos 2020/21 ($P_1$-$P_6$) y 2019/20 ($P_1$-$P_5$).*

| V | $t_d$ | $t_c$ | Contenido | P |
|---|---|---|---|---|
| $V_0$ | 2' 4'' | - | - | - |
| $V_1$ | 3' 5'' | 30' | Qué es | - |
| $V_2$ | 4' 9'' | 30' | Cómo se puede mover (traslación y rotación) | $P_1$ |
| $V_3$ | 3' 17'' | 20' | Cómo componer traslaciones | |
| $V_4$ | 3' 20'' | 3' 5'' | Cómo componer rotaciones | |
| $V_5$ | 4' 6'' | 30' | Cómo se compone un par de rotaciones | |
| $V_6$ | 1' 19'' | - | Cómo se componen traslaciones y rotaciones | |
| $V_7$ | 3' 31'' | - | Cómo se relaciona la velocidad de un punto con la de otro | $P_2$ |
| $V_8$ | 4' 11'' | 60' | Cuáles son los invariantes | $P_3$ (y $P_6$) |
| $V_9$ | 4' 14'' | - | Cómo se calcula el eje instantáneo de rotación y deslizamiento | $P_4$ y $P_5$ |

Con el objetivo de aumentar la motivación de los alumnos e incorporar ciertas relaciones Ciencia-Tecnología-Sociedad, en el ejercicio correspondiente al vídeo $V_6$ se planteaba a los alumnos que visualizaran un par de vídeos dedicados al parkour y al "skate", y que dibujaran las velocidades de traslación y rotación en dos momentos concretos.

Este curso, la asignatura ha contado con 46 alumnos matriculados, de los que 38 eran de nuevo ingreso. Para llevar a cabo la evaluación, después de visionar cada vídeo, se pedía a los alumnos que cumplimentaran un breve cuestionario con entre dos y diez apartados a través de Moodle, tanto con limitación de tiempo como sin ella (ver Tabla 1). Para asegurarnos un correcto proceso de enseñanza-aprendizaje, una vez pasado el periodo de visualización de los vídeos, se llevó a cabo una sesión on-line de 1 hora de duración dedicada a profundizar en los conceptos más abstractos. En esta sesión, el profesor clarificó, por medio de la lección magistral, algunos conceptos a los alumnos como qué significa un par de rotaciones o qué es un eje instantáneo de rotación y deslizamiento; hay que indicar que este último concepto requiere de una exposición relativamente larga, frente a las explicaciones breves (y mucho más sencillas) de los vídeos usados en las tareas en casa. Una vez impartida toda la teoría hubo dos sesiones de 1 hora y 15min cada una en las que, primero el profesor y luego los alumnos bajo su supervisión, resolvieron problemas más complejos (similares a los del examen), que combinaban todos los conceptos tratados en los vídeos y test previamente realizados. Durante las sesiones anteriores, así como a lo largo de las tres semanas que duró la experiencia, el profesor también resolvió las dudas que tenían los alumnos. Además, los alumnos que lo deseasen podían realizar dos ejercicios optativos para subir nota (similares a los explicados en las últimas dos sesiones con el profesor). Por último, se dio una semana de plazo para que los alumnos que lo quisieran repitiesen los cuestionarios (una vez conocida la calificación de los mismos) o los hiciesen por primera vez (en caso de no haberlos completado antes) después de las tres sesiones on-line. De esta forma, nos asegurábamos de que la evaluación se llevaba a cabo después de que todos los alumnos hubieran tenido la oportunidad de resolver sus dudas y corregido posibles errores, así como haber recibido la retroalimentación necesaria. Además, nos asegurábamos de que podían enmendar, los problemas que podían haber tenido a la hora de responder a los cuestionarios. Por último, se pidió a los alumnos que respondieran a una encuesta de satisfacción con 15 preguntas para valorar la metodología, el tiempo y materiales utilizados y posibles propuestas de mejora.

Para comprobar la efectividad de la metodología utilizada, hemos tomado como referencia dos tipos de resultados. Por un lado, hemos comparado las calificaciones en la evaluación continua de CSR con las de cada uno de los ejercicios de los temas 1-4 de los exámenes parcial (celebrado en noviembre de 2020) y final (febrero de 2021) de este curso. Hay que indicar que el examen parcial tiene carácter eliminatorio, por lo que los alumnos que lo aprueban (calificación mayor o igual que 5,0) no tienen que examinarse de esa parte en el final. La Tabla 1 establece cómo se relacionan las diferentes preguntas $P_1$-$P_6$ del ejercicio de CSR con los vídeos empleados. No se han considerado las notas de los ejercicios correspondientes a los temas 5 y 6 porque éstos sólo se evalúan en el examen final. La comparativa de las calificaciones anteriores permite determinar la efectividad del aula invertida en comparación con la docencia on-line *más* tradicional usada en los otros temas, que se ha basado en lecciones magistrales, resolución de problemas, prácticas de laboratorio, etc.

Por otro lado, hemos estudiado, también, las calificaciones en los exámenes parcial y final del pasado curso 2019/20 para identificar posibles sesgos debido al particular desarrollo de este curso (docencia on-line, poco contacto con los compañeros, posibles situaciones de ansiedad y estrés de los alumnos, etc.). Esto nos permite contextualizar el aprendizaje del tema de CSR respecto a los demás, evitando las posibles distorsiones debido al uso de una docencia on-line. Hay que indicar que el curso pasado el ejercicio de CSR tenía un apartado menos que el actual, por lo que únicamente constaba de 5 preguntas (no aparecía la pregunta $P_6$ de cálculo de la velocidad mínima del SR).

El análisis de datos mencionado se ha realizado introduciendo las matrices con las calificaciones de los alumnos en los ejercicios en distintas hojas de cálculo MS Excel. Para llevar a cabo una correcta comparativa de ejercicios individuales, tods calificaciones se han normalizado a 10.

4. RESULTADOS

Como se ha comentado en la sección anterior, después de visualizar cada uno de los vídeos del tema de CSR, los alumnos debían responder un breve cuestionario. Pese a que se tenía en cuenta para la evaluación continua, la mayoría de los alumnos no realizaron todos los test con los que se evaluaba el aprendizaje del tema únicamente los hizo todos uno de los estudiantes). En total, hubo 25 alumnos (54% del total) que participaron en algún momento en la experiencia del aula invertida. No obstante, la media de participantes en la experiencia fue de unos 13 alumnos (28% del total), como se recoge en la Tabla 2, de lo que sólo uno era repetidor (que aprobó el examen). Al comparar las calificaciones de los alumnos y relacionarlas con la duración de los vídeos y el tiempo que se tenía para responder a los cuestionarios, podemos concluir que ninguno de estos dos últimos tienen influencia significativa.

*Tabla 2: Número de alumnos (N) que responden a los cuestionarios usados para evaluar el aprendizaje alcanzado con el aula invertida tras la visualización de los vídeos $V_1$-$V_9$ durante el curso 2020/21, y calificaciones medias ($\mu$) y desviaciones estándar ($\sigma$) obtenidas. El número entre paréntesis (N') indica el número de alumnos que hacen el test después de las 3 sesiones on-line con el profesor (repitiéndolo o haciéndolo por primera vez).*

| V | $V_1$ | $V_2$ | $V_3$ | $V_4$ | $V_5$ | $V_6$ | $V_7$ | $V_8$ | $V_9$ |
|---|---|---|---|---|---|---|---|---|---|
| N (N') | 20 (1) | 15 (0) | 15 (3) | 16 (1) | 10 (0) | 5 (0) | 19 (2) | 10 (2) | 9 (3) |
| $\mu$ | 7,8 | 8,2 | 7,4 | 2,8 | 9,0 | 10,0 | 6,3 | 3,8 | 2,7 |
| $\sigma$ | 1,8 | 2,0 | 4,0 | 3,6 | 2,1 | 0,0 | 2,7 | 2,9 | 2,2 |

En primer lugar, hay que destacar la alta dispersión que presentan los datos, que tiene, en general, una desviación típica con valor mayor o igual a 2,0. Esto se podría deber a la disparidad en la formación académica de los alumnos, por lo que las conclusiones que se detallan a continuación respecto a las calificaciones obtenidas por los alumnos en los cuestionarios de los vídeos deben tomarse con cierta cautela.

Como cabía esperar, los vídeos $V_1$ y $V_2$ han sido respondidos por un número considerable de alumnos obteniendo una nota media relativamente elevada (alrededor de 8,0 de media), debido a que trataban los conceptos más básicos y elementales de CSR. Los resultados son también bastante buenos para la composición de traslaciones ($V_3$), al tratarse de algo relativamente sencillo, pero son considerablemente peores en cuanto a la composición de rotaciones ($V_4$), ya que requiere un mayor nivel de abstracción. Sorprenden, por ello, los excelentes resultados obtenidos para el vídeo $V_5$, que se refiere al cálculo de la velocidad producida por un par de rotaciones.

De entre todos los vídeos, destacan, también, los excelentes resultados obtenidos para el vídeo $V_6$, en el que describe cómo calcular la velocidad de un punto de un SR en función de la de otro. Aunque el cuestionario correspondiente es el menos respondido (tan sólo lo completan 5 alumnos), todos ellos lo hacen sin errores. Por otro lado, sobresalen también las respuestas en el vídeo $V_7$. En este caso, los alumnos tenían que realizar una tarea asociada a la composición de traslaciones y rotaciones con una fuerte componente CTS relacionada, además, con cuestiones de interés de los alumnos, por lo que fue respondida por un gran número de ellos (21). Los resultados referidos a los dos últimos vídeos son bastante peores debido a que se referían a los dos conceptos más complejos del tema CSR: el cálculo de los invariantes ($V_8$) y del eje instantáneo de rotación y deslizamiento ($V_9$) de un SR.

Por último, hay que indicar que han sido pocos los alumnos que han realizado las actividades para subir nota, tanto la repetición de los cuestionarios, como los problemas extra propuestos. Estos últimos solo han sido entregados por tres alumnos, obteniendo unas calificaciones extremadamente bajas (de 0, 1 y 2 puntos sobre 10). Estos hechos parecen poner en tela de juicio la eficacia del proceso enseñanza-aprendizaje que, a juzgar por los resultados recogidos en la Tabla 2, invitaban en un principio al optimismo.

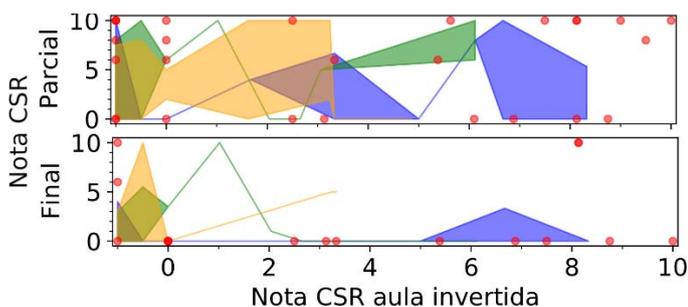

*Figura 1: Calificación de los apartados del ejercicio de cinemática del sólido rígido en los exámenes parcial (arriba) y final (abajo) del curso 2020/2021 frente a la calificación obtenida en los cuestionarios de aula invertida asociados a dichos apartados (puntos rojos: nota en la pregunta $P_1$; áreas azul, naranja y verde: límites de las notas en las preguntas $P_2$, $P_3+P_6$ y $P_4+P_5$).*

Con el objetivo de analizar en más profundad la efectividad del aula invertida en el proceso de enseñanza-aprendizaje, en la Figura 1 mostramos las respuestas obtenidas en cada uno de los apartados del ejercicio de CSR de los exámenes parcial (arriba) y final (abajo) de este curso 2020/21 en función de las correspondientes calificaciones en los cuestionarios usados en el aula invertida. Para distinguir las respuestas asociadas a vídeos no visualizados de las de las respondidas de manera errónea, se ha decidido asignar calificaciones negativas a aquellos alumnos que no han respondido todos los cuestionarios y, por tanto, asumimos que tampoco han visualizado todos los vídeos. En particular, se ha calificado con -0,5 los cuestionarios puntualmente no respondidos y con -1 si los cuestionarios no se han respondido en general.

Hay que indicar que, al no coincidir exactamente los cuestionarios de los vídeos con las preguntas del examen, que en general eran más complejas al involucrar varios conceptos, ha sido necesario usar la relación contenida en la Tabla 1. Debido a la alta dispersión que tienen los datos, para visualizar e interpretar mejor los resultados, hemos decidido presentar una sola serie de datos con puntos y mostrando los valores de las demás en áreas sombreadas que acotan la región en la que se encuentran. En el caso particular de la Figura 1, hemos representado las calificaciones para la pregunta $P_1$. Como se puede observar, los datos presentan una gran dispersión, debida, en buena medida, a la dispar formación de los estudiantes. Los datos no permiten afirmar que aquellos estudiantes con buenas calificaciones en la evaluación de aula invertida obtengan, en general, mejores notas en los apartados correspondientes en el examen. Además, hay que tener en cuenta que aproximadamente un 50% de los alumnos, no ha contestado a la mitad de los cuestionarios de los vídeos, por lo tanto, no se puede establecer una relación con los resultados del examen. No obstante, los datos demuestran que, en general, los estudiantes que sacan buenas calificaciones en la evaluación del aula invertida lo hacen también en el apartado correspondientes del ejercicio del examen.

Para establecer qué importancia tienen los distintos apartados del ejercicio de CSR sobre la calificación global del mismo, en la Figura 2 mostramos esta última en función de las calificaciones obtenidas en los distintos apartados $P_1$-$P_6$ para los exámenes parcial (arriba) y final (abajo). De nuevo, se puede apreciar una gran dispersión en los datos Por ello, para determinar cómo contribuyen los distintos apartados a la nota del ejercicio, presentamos en la Tabla 3 el coeficiente de correlación de Pearson que, como se puede observar, es en todos los casos positivo y relativamente elevado, lo que demuestra la importancia de todos ellos en la calificación global del ejercicio CSR.

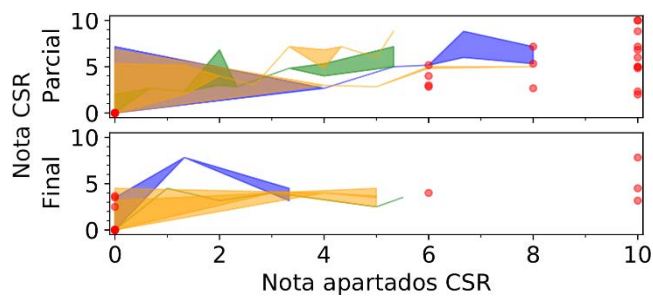

*Figura 2: Calificación global del ejercicio de cinemática del sólido rígido en los exámenes parcial (arriba) y final (abajo) del curso 2020/21 frente a la calificación global obtenida en cada uno de sus apartados (puntos rojos: nota en la pregunta $P_1$; áreas azul, naranja y verde: límites de las notas en las preguntas $P_2$, $P_3+P_6$ y $P_4+P_5$).*

Vista la influencia de los diferentes apartados, vamos a estudiar los resultados obtenidos por aquellos alumnos que han seguido aula invertida y los que no y así ver si hay una influencia real en la mejora del rendimiento académico. En el examen parcial, la media obtenida por los alumnos que llevaron a cabo aula invertida fue de 3,94 y la de aquellos que no visionaron los videos fue de 2,76. Tras la realización de las pruebas estadísticas correspondientes (Test F y Test T), se obtiene que no existen diferencias significativas entre las medias, por lo que, aunque la nota sea considerablemente más alta, no se puede afirmar que el empleo de aula invertida suponga una mejora en la nota del ejercicio en cuestión.

*Tabla 3: Coeficiente correlación de Pearson entre la nota de los apartados $P_1$-$P_{5/6}$ y la nota los exámenes parcial y final de los cursos 2020/21 y 2019/20.*

| P | 2020/21 | | 2019/20 | |
|---|---|---|---|---|
| | Parcial | Final | Parcial | Final |
| $P_1$ | 0,8865 | 0,7982 | 0,7171 | 0,7130 |
| $P_2$ | 0,8251 | 0,6220 | 0,8043 | 0,7268 |
| $P_3$ | 0,9225 | 0,7989 | 0,8482 | 0,8357 |
| $P_4$ | 0,7218 | 0,6334 | 0,8553 | 0,8316 |
| $P_5$ | 0,8758 | 0,8316 | 0,7330 | 0,6627 |
| $P_6$ | 0,8359 | 0,8465 | - | - |

En lo que respecta al examen final, de nuevo las diferencias entre las medias obtenidas (1,47 vs 2,39) no presentan *a priori* diferencias estadísticamente significativas. No obstante, en este caso, hay que tener en cuenta también que el número de alumnos que se presentan al examen y que no han llevado a cabo aula invertida es únicamente de 3 (frente a los 15 que sí la usaron).

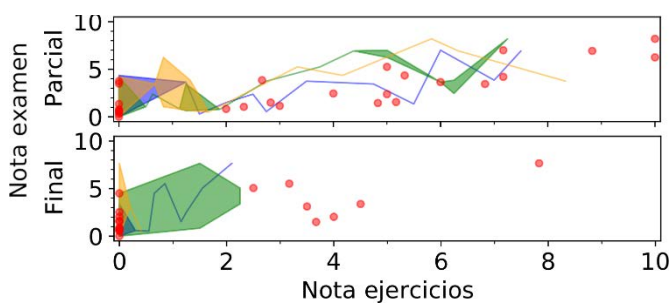

*Figura 3: Calificación global de los exámenes parcial (arriba) y final (abajo) del curso 2020/2021 frente a la calificación obtenida en cada uno de los ejercicios de dichos exámenes (puntos rojos: nota en cinemática del sólido Rígido; áreas azul, verde y naranja: límites de las notas en las preguntas de cinemática del punto 1, cinemática del punto 2 y movimiento relativo).*

Para concluir la discusión sobre los resultados del aula invertida, en la Figura 3 mostramos la calificación global de los exámenes parcial (arriba) y final (abajo) del curso 2020/21 en función de las obtenidas en los 4 ejercicios que formaban la primera parte del temario (dos de cinemática del punto, el de CSR y otro de movimiento relativo). De nuevo, los datos presentan una dispersión considerable. No obstante, se observa en todos ellos la misma tendencia: en general los alumnos que sacan mejores calificaciones en cada uno de los ejercicios obtienen mejor calificación global en el examen. La nota media en el ejercicio de CSR es la más alta de todos (3,6 puntos en el parcial y 1,6 puntos en el final, frente a 3,1 y 1,0 puntos sacados en el siguiente ejercicio con más puntuación, respectivamente). Esta mayor nota media se puede afirmar también, observando que hay más puntos por debajo de la línea y=x para el ejercicio de CSR que para el resto. Sin embargo, esta diferencia no es estadísticamente significativa, debido al gran valor que toma la desviación estándar. Por otro lado, el ejercicio de CSR es el más correlacionado con la nota global de los exámenes (el parámetro de regresión es $R^2 = 0,64$ y 0,51 en los exámenes parcial y final, frente a $R^2 \approx 0,52$ y 0,27 para los de cinemática del punto, respectivamente). En general, el parámetro de regresión es superior en los ejercicios del examen parcial que para los del final, debido a que éste último tenía más ejercicios relacionados con otras partes del temario. Las menores calificaciones en el examen final frente al parcial se pueden explicar debido a que los ejercicios de este último examen únicamente los tenían que realizar los alumnos que no habían aprobado el parcial.

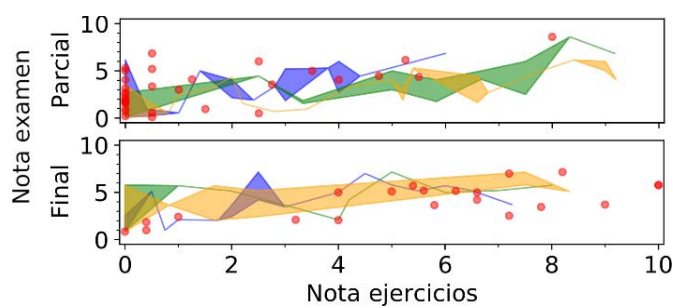

*Figura 4: Calificación global de los exámenes parcial (arriba) y final (abajo) del curso 2019/2020 frente a la calificación obtenida en cada uno de los ejercicios de dichos exámenes (puntos rojos: nota en cinemática del sólido rígido; áreas azul, verde y naranja: límites de las notas en las preguntas de cinemática del punto, dinámica del punto y movimiento relativo).*

Por último, con el objetivo de contextualizar el tema de CSR, mostramos en la Figura 4 las calificaciones de los exámenes parcial (arriba) y final (abajo) del pasado curso 2019/20. En este caso, los exámenes constaban, también, de cuatro ejercicios, pero en lugar de tener dos de cinemática del punto tenían uno de este tema y otro de dinámica del punto. Al igual que en el curso 2020/21, la calificación media obtenida en el curso 2019/20 en el ejercicio de CSR es mayor que en el resto de los ejercicios (3,5 en el parcial y 5,4 en el final frente a los 1,9 y 2,7 puntos obtenidos, respectivamente, en el movimiento relativo, que ha sido el segundo con mayor nota). Sin embargo, como en el caso anterior, estos resultados deben tomarse con cautela dada la elevada dispersión que tienen (superior a 2,0). Por último, el parámetro de regresión toma en el ejercicio de CSR valores similares a los de este curso ($R^2 = 0,57$ y 0,52), y es mayor al del resto de los ejercicios en todos los casos salvo en el del ejercicio de movimiento relativo del examen parcial ($R^2 = 0,65$). La elevada dispersión estadística de los resultados hace imposible concluir de forma precisa cuál ha sido el impacto del aula invertida en el proceso de enseñanza-aprendizaje, pero los resultados obtenidos demuestran que su impacto ha sido moderado y no especialmente significativo. Hay que tener en cuenta, además, que la docencia durante el curso 2020/2021 ha sido on-line (salvo las dos sesiones presenciales de prácticas de laboratorio) lo cual influye mucho en el buen funcionamiento y el aprovechamiento de las sesiones de aula llevadas a cabo tras el visionado de los vídeos.

En lo que respecta a la satisfacción de la experiencia por parte de los alumnos, se ha observado que los alumnos valoran positivamente el aula invertida (calificación media de 6,2) y el aprovechamiento de las sesiones posteriores a la visualización de los vídeos (calificación media de 6,4). Este último resultado está en concordancia con las respuestas respecto a su preferencia por enseñanza más tradicional. El promedio de horas invertidas en el aula invertida ha sido de 5h y sólo el 27% de los alumnos considera que invierte más tiempo usando esta metodología. En cuanto al material, todos han usado ordenador para la visualización de los vídeos y el 73% de ellos han visionado los vídeos más de una vez. Sólo 3 de los 11 alumnos encuestados han recurrido a otros recursos para comprender los contenidos. En general, en las opiniones mostradas los alumnos echan de menos resolver sus dudas de manera inmediata, pero valoran positivamente poder ver los vídeos más de una vez.

En un marco más amplio, se ha observado un considerable deterioro de las tasas globales de la asignatura, probablemente debido al extraño entorno en el que se ha llevado la docencia este curso 2020/21. Así, tanto la tasa de rendimiento como la de éxito se han reducido en casi un 50%, (han pasado de un 32,1% y un 56,7% el curso 2019/20 a tan sólo un 15,9% y 32,1% este curso 2020/21). La tasa de absentismo también se ha deteriorado, aunque de forma más moderada al pasar de un 43,4% el curso pasado a un 50,0% éste.

Hay que indicar que el resto de las asignaturas impartidas durante el mismo periodo han presentado un comportamiento similar.

5. CONCLUSIONES

En este trabajo hemos presentado una experiencia docente on-line con estudiantes universitarios de Física I (primer curso del Grado en Ingeniería Agroambiental de la Universidad Politécnica de Madrid) basada en la aplicación del aula invertida para el estudio de cinemática de sólido rígido. En vista a los resultados mostrados, no se puede afirmar que el empleo de aula invertida tenga un efecto positivo en el rendimiento académico del grupo de estudiantes estudiado, en contra de lo esperado por estudios previos en la literatura. Este hecho se puede atribuir a las especiales circunstancias en las que se ha llevado a cabo la docencia por la situación ocasionada por la pandemia de COVID-19 durante este curso 2020/2021 (docencia on-line, poco contacto entre los alumnos, etc.). No obstante, se ha observado que los alumnos valoran positivamente el uso de esta metodología, especialmente por el mejor aprovechamiento de las sesiones de clase, aunque sienten preferencia por la docencia más tradicional.